\documentclass[10pt,twocolumn]{article}
\usepackage{times}

\pdfoutput=1

\usepackage{amsmath}
\usepackage{amssymb}
\usepackage[numbers]{natbib}
\usepackage{booktabs} 
\usepackage{enumitem} 
\usepackage{soul} 
\usepackage{url} 
\usepackage{hyperref}
\usepackage{xcolor}

\usepackage[newfloat,frozencache]{minted}

\usepackage{mdframed}

\usepackage{todonotes}

\usepackage{subcaption}

\begin{document}


\title{\bf The Case for MUSIC: A Programmable IoT Framework for Mobile
  Urban Sensing Applications}


\author{Shiva R. Iyer$^1$, Soumie Kumar$^1$, Kate Boxer$^2$, Fatima Zarinni$^1$, \\
  Lakshminarayanan Subramanian$^1$ \\
  \small {\em $^1$Department of Computer Science, Courant Institute of Mathematical Sciences, New York University} \\
  \small {\em $^2$Department of Computer Science, Columbia University} \\
  \small Submission Type: Vision
}

\date{}

\maketitle

\begin{abstract}

  This vision paper presents the case for MUSIC, a programmable
  framework for building distributed mobile IoT applications for urban
  sensing.  The Mobile Urban Sensing, Inference and Control (MUSIC)
  framework is contextualized for scenarios where a distributed
  collection of static or mobile sensors collectively achieve an urban
  sensing task.  The MUSIC platform is designed for urban-centric sensing applications 
  such as location sensing on mobile phones for road traffic
  monitoring, air quality sensing and urban quality monitoring using
  remote cameras.  This programmable system, at a high level, consists
  of several small sensors placed throughout a city on mobile vehicles
  and a centralized controller that makes decisions on sensing in
  order to achieve certain well-defined objectives such as improving
  spatial coverage of sensing and detection of hotspots.  The system
  is programmable in that our framework allows one to create custom
  smart systems by writing custom control logic for sensing. Our
  contributions are two-fold -- a backend software stack to enable
  centralized control of distributed devices and programmability, and
  algorithms for intelligent control in the presence of practical
  power and network constraints. We briefly present three different
  urban sensing applications built on top of the MUSIC stack.
\end{abstract}


\vspace{-2mm}
\section{Introduction}
\label{sec:intro}
\vspace{-2mm}

The vision of {\em smart cities} is being largely powered by a broad
array of new wireless telemetry applications for addressing urban
challenges.  Urban-centric wireless telemetry applications such as air
quality sensing
\cite{Bigazzi2015,Dong2015,Shi2016,Devarakonda2013Realtime}, road
quality monitoring \cite{Eriksson2008Pothole}, road traffic delay
estimation \cite{Thiagarajan2009Vtrack} and fleet tracking
\cite{Samsara} rely on a large number of {\em mobile IoT} sensors that
are controlled by a cloud controller to collectively achieve a
distributed urban sensing task. Similar telemetry applications in
other contexts include: smart agriculture
\cite{Kim2008Remote,Goumopoulos2014Automated,Ahmed2011Innovative,Xiao2010Smart},
smart water networks \cite{Kartakis2015Waterbox}, wildlife monitoring
\cite{Kumar2015Zigbee}, and many others \cite{Xiaoli2011Remote}.

For many of these applications, especially those in that are
applicable in urban contexts, fine-grained sensing is essential
thereby placing a minimum need on the number of sensors. For example,
a city wide deployment of a fine-grained air quality monitoring system
may require 500-1000 sensors.\footnote{Extrapolating based on our
  experiences deploying a small-scale air quality monitoring system in
  Delhi over a 30 km$^2$ area.}  Most urban sensing applications may
also require a high frequency of sensing (such as 1 Hz or higher)
especially at certain times such as when the quantity being measured
is changing rapidly at real time. Applications such as pothole
detection and monitoring using accelerometer
tracking\cite{Eriksson2008Pothole} or location tracking using cellular
network signals \cite{Thiagarajan2011Accurate} may require even higher
sampling rates.

To manage a large scale of mobile IoT sensors for these applications
in an efficient manner, we need smart sensing and control policies to
control the granularity of sensing, the granularity of data collection
and the granularity of computations. These mobile IoT applications
also have a significant impact on the cellular network footprint and
the power consumption by the tiny mobile devices. According to
\cite{Cisco2018}, the amount of useful data generated by IoT devices
in 2021 will be about 85 ZB, but only 7.2 ZB of that will actually be
stored or used. Our experiments with air quality monitoring for this
work have also shown that it costs approx 30 MB of data per sensor per
day in order to obtain fine-grained air quality data at the rate of 1
Hz. For a city-wide deployment, this amounts to at least 15 GB of data
per day (which also incurs non-trivial network data costs with a
yearly opex more than the capex of deployment). Power consumption by
mobile devices are also a problem, which could be reduced if the
sensors were turned off when not needed or if the sensors generated
only the amount of data that is actually needed.




This paper describes the design of the Mobile Urban Sensing Inference
and Control (MUSIC) stack that aims to address many of these
challenges outlined above. The MUSIC platform is designed to enable
easy development of cloud controlled distributed mobile sensing
applications. In MUSIC, a cloud controller can control a distributed
collection of mobile devices and sensors using a type of \emph{sensing
  policy} and objective that can be easily specified.  In this paper,
we describe the MUSIC system, consisting of the stack built on top of
conventional networking stacks, and show how it can be used in
providing flexibility and programmability in three real world sensing
applications -- spatial coverage mapping for air quality sensing,
hotspot detection for road traffic analysis and dirt detection for
urban cleanliness monitoring.  The MUSIC platform is \emph{flexible}
to work with a diverse array of sensors that can connect to a mobile
device, possibly from multiple vendors, and \emph{programmable} to
support custom sensing strategies given an objective sensing function
and constraints. The MUSIC platform can deal with three constraints:
(a) power-awareness of edge devices; (b) cost-awareness due to network
costs that may vary by country and application needs; (c) network awareness
to make the application resilient in the face of intermittent connectivity
and lack of reliable sensing data.


In this vision paper, we specifically aim to demonstrate the case for
a MUSIC stack and demonstrate the utility of the stack using real
world urban sensing applications. We have early experiences tailoring
the MUSIC platform for three applications: air quality monitoring in
Delhi, traffic hotspot detection in New York City and road cleanliness
detection in Delhi.  All these applications have been built using a
cloud service and an Android mobile application which can interconnect
with in-built, fixed or bluetooth sensors.





\vspace{-2mm}
\section{Related Work}
\label{sec:intro-related}
\vspace{-2mm}

There have been a broad array of works on mobile IoT applications,
urban sensing applications, wireless telemetry and specialized 
IoT applications. However, we are unaware of any
generic and programmable platform that can support several 
distributed sensing applications with flexible sensing
policies and programmable objective functions and constraints.
We outline some of the key related works.

\textbf{Smart sensing for specialized applications:} In the field of
agriculture, there have been works on making irrigation more smart in
order to reduce wastage of water. In \cite{Kim2008Remote} and
\cite{Goumopoulos2014Automated}, the authors implement WSNs for remote
sprinklers, along with sensing and control for variable rate
irrigation. \cite{Xiao2010Smart} describes a full-fledged system for
smart irrigation where the backend server takes real-time moisture
data as input to determine sensing decisions. In the context of urban
air quality monitoring, given increasing concern over poor air quality
in many parts of the world, there has been a recent surge of
commercially available low-cost portable sensors by various vendors
\cite{aircasting,airveda,kaiterra}. The drawback is that each vendor
has its own frontend and backend system that is usually compatible
only with its own products. This becomes a challenge when conducting
large-scale urban air quality monitoring because it restricts freedom
in choosing devices based on its specifications. Likewise, there have
also been numerous experimental works on mobile urban air quality
monitoring
\cite{Dutta2009CommonSense,Devarakonda2013Realtime,Bigazzi2015,Dong2015,Shi2016}. While
all of them contain real-time reporting in some form, we have not
found evidence for programmability and extensibility for implementing
custom policies.

\textbf{Similar Architectures:} Mobile Fog \cite{Hong2013Mobile} is a
system developed for IoT applications such as vehicle tracking using
cameras and traffic monitoring MCEP. While they semonstrate the
ability to perform analytics at reduced latency and network bandwidth,
there is no specific focus on enabling a flexible programmable
stack across applications. Ravel \cite{Riliskis2015Ravel} is a system
designed to aid ease of development of IoT applications but using the
traditional MVC architecture, which calls for user input for control.

\vspace{-2mm}
\section{The MUSIC Programmable Stack}
\label{sec:system}
\vspace{-2mm}


Whereas most smart sensing systems today have available hardcoded
policies in them, we envision the ability to program custom policies
in the MUSIC platform. The policies would serve primarily to control
dynamically the amount of data that is being collected by the sensors
on the ground via three means -- turning specific sensors off when not
needed, varying the sensing or reporting frequency and suggest changes
in the spatial position of the sensor. This type of control requires a
centralized ``controller'' that monitors the data from all the sensors 
and provides continuous feedback.

\begin{figure}[h]
  \centering
  \includegraphics[scale=0.3]{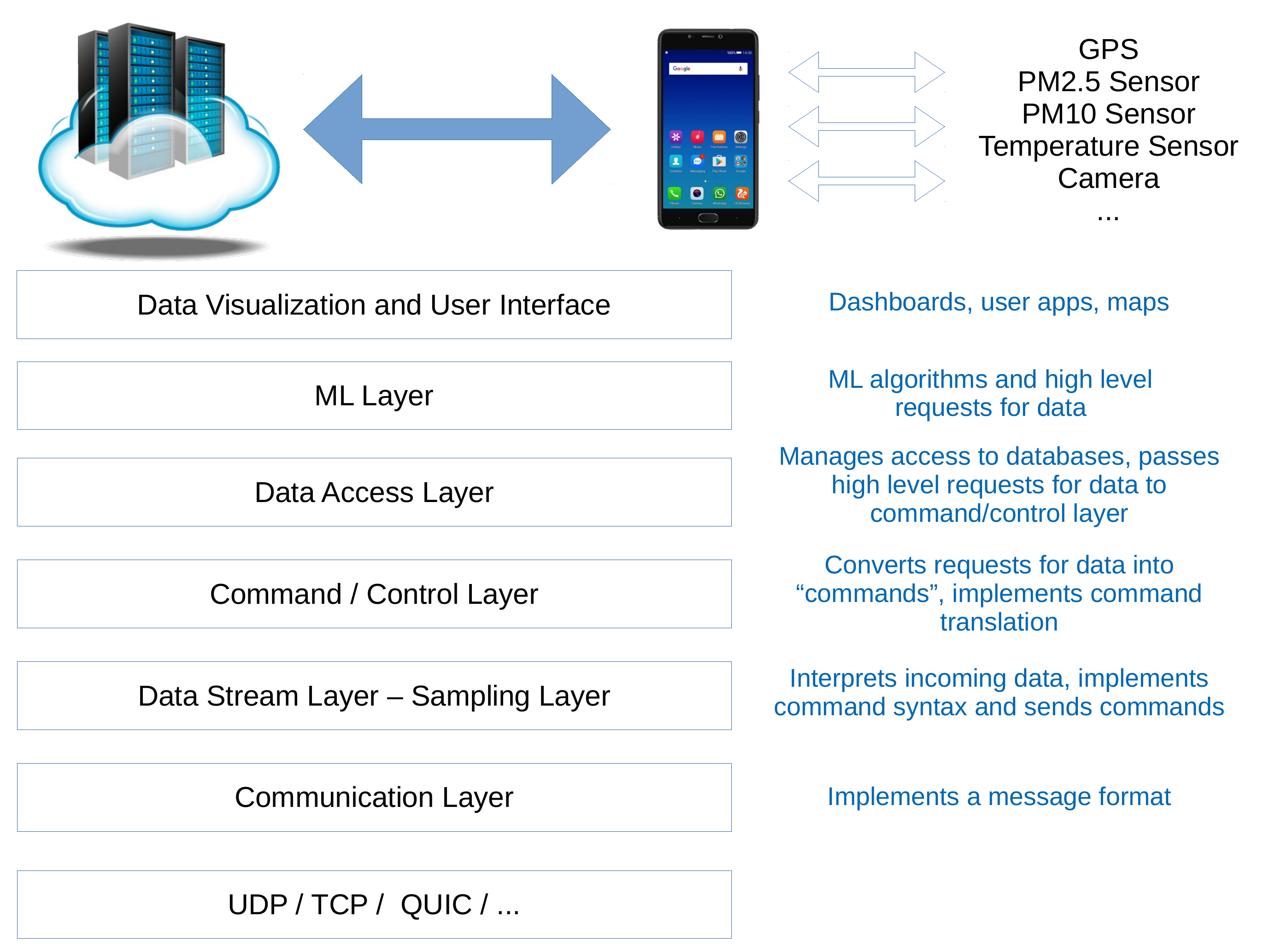}
  \caption{Layers of the MUSIC Stack}
  \label{fig:music-layers}
\end{figure}

The MUSIC platform is a modular layered stack built over regular
communication channels to enable the intelligent and ML-based
control. Figure~\ref{fig:music-layers} illustrates the stack.  The
highest layer is responsible for data analytics and visualization
based on collected sensing data. The vision of the ML control layer is
to support different machine learning and standard optimization
algorithms to help in determining the best policies to order to
minimize our network, cost and power constraints, translates the
policies into commands. The command control layer translates these
commands to actionable commands to individual devices over existing
communication data channels which is maintained by the data stream
layer.



MUSIC is an end-to-end system that consists of a mobile phone
application and a backend server-cum-controller. The frontend mobile
app is the gateway for sensors to report data. Sensors are either
built-in ones in the phone (e.g. accelerometer, GPS or compass) or are
external ones that communicate over Bluetooth (e.g. air quality
monitors).  We emphasize that the gateway does not need to be a mobile
phone, but could be a standalone sensor with wireless connection
capabilities and general purpose computing hardware. The backend
serves to provide access to the data as well as implement control by
sending ``commands'' to the mobile phones, such as START and STOP,
based on sensing decisions.  The mobile app is adapted from an
existing open source project \cite{aircasting} and we used it for air
quality monitoring as well as collecting GPS traces, camera and
accelerometer readings for traffic and road cleanliness applications.
Next, we outline the key properties of many of these layers.




\vspace{-1.5mm}
\subsubsection*{Data Layer}
\label{sec:data-layer}
\vspace{-1.5mm}

Built over the transport layer, either TCP or UDP, this layer sets the
format for data that is sent from the IoT devices to the cloud. There
is a master \emph{Driver} thread, that has a server-side TCP socket
opened for incoming connection requests from edge nodes. A new thread
is spawned for each new edge node that is connected, called the
\emph{phone thread}, in order to receive data from the node as well as
to send control commands.

There are broadly two types of messages sent by the edge nodes -- a
\emph{KeepAlive} message and a data message. The former is a periodic
\emph{ping}-style message that is aimed at informing the backend about
the device's presence as well as the changing IP address of the mobile
edge. It contains fields such as time stamp, latitude, longitude and
battery life. The full set of fields in a KeepAlive message is shown
in listing \ref{lst:keepalive-json}. The second type of message is a
data message, which contains data collected by the sensors. The exact
set fields in this message depends on the sensor data that is
reported. We show two listings below (\ref{lst:image-json} and
\ref{lst:data-json}), for camera image data and other sensor data
respectively.

\begin{listing}[ht]
  \begin{minted}[fontsize=\scriptsize]{json}
    {
      "battery_life" : 95,
      "imei" : "353323062860043",
      "ip" : "172.16.19.89",
      "isData" : false,
      "isImage" : false,
      "keep_alive_status" : true,
      "latitude" : 40.7348562,
      "longitude" : -73.9949165,
      "sensors" : ["Accelerometer", "Compass"]
    }
  \end{minted}
  \caption{Fields in KeepAlive message}
  \label{lst:keepalive-json}
\end{listing}

\begin{listing}[ht]
  \begin{minted}[fontsize=\scriptsize]{json}
    {
      "imei" : "353323062860043",
      "isData" : false,
      "isImage" : true,
      "latitude" : 40.7348562,
      "longitude" : -73.9949165,
      "encodedImageString" : "..."
    }
  \end{minted}
  \caption{Fields in the JSON (for image data)}
  \label{lst:image-json}
\end{listing}

\begin{listing}[ht]
  \begin{minted}[fontsize=\scriptsize]{json}
    {
      "imei" : "353323062860043",
      "isData" : true,
      "isImage" : false,
      "latitude" : 40.7348562,
      "longitude" : -73.9949165,
      "sensorData" : {
        "accelerometer" : [{
          "sensorName" : "accelerometer",
          "sensorSessionData" : [{
            "measurement_unit" : 
            "meters per second squared",
            "name" : "Accelerometer",
            "x" : -0.15322891,
            "y" : 3.7828386,
            "z" : 8.820239
          }]
        }]
      }
    }
  \end{minted}
  \caption{Fields in the JSON (for other sensor data)}
  \label{lst:data-json}
\end{listing}

\vspace{-1.5mm}
\subsubsection*{Sampling Layer for Control Communication}
\label{sec:control-layer}
\vspace{-1.5mm}

We refer to this as the ``sampling'' or ``control'' layer. As figure
\ref{fig:music-layers} shows, the sampling layer is built over a
transport layer protocol such as TCP. In our implementation thus far,
we have used TCP, even though it can be implemented equally over UDP
as well. Mobile devices keep separate TCP connections to the cloud for
this purpose for the cloud to send commands to the device, such as
START and STOP. These commands perform the simple tasks as the names
indicate -- a START command received by the edge (mobile phone)
results in the app starting a new recording session. The particular
sensors to be started, as well as the frequency at which the sensors
are to report data, are provided as arguments to the command. The STOP
likewise causes the current recording session to end. The data
collected between a START and a STOP is saved temporarily in the phone
until a SEND command is received, at which point the data is sent to
the server. Whether the data is compressed or not is indicated by an
argument to the command. The message is a simple text JSON as shown in
listing \ref{lst:cmd-msgs}.

\begin{listing}[ht]
  \begin{minted}[fontsize=\scriptsize]{json}
    {"messageType" : "START",
      "sensor" : "",
      "frequency" : ...}
    {"messageType" : "STOP"}
    {"messageType" : "SEND"}
  \end{minted}
  \caption{Command messages sent by sampling layer}
  \label{lst:cmd-msgs}
\end{listing}

As the communication happens over a cellular network, the IP address
associated with the mobile edge would not remain fixed, and therefore
the phone periodically sends \emph{ping} packets to the server, which
serves two purposes -- i) it helps to know if the edge is alive as the
lack of such packets most probably indicate an edge that died or that
is malfunctioning, ii) the server is constantly updated about the IP
address of the edge node to send commands to. This type of control is
akin to the control in software-defined networks, where the central
controller writes rules in the flow tables in the routers. In our
application, mobile phones are the nodes that act as middleboxes and
the cloud controller sends commands to them. The commands we support
are shown in listing \ref{lst:commands}.

\begin{listing}[ht]
  \begin{minted}[fontsize=\scriptsize]{java}
    static enum Command {
      START,
      STOP,
      SEND,
      CAPTURE_IMAGE;
    }
  \end{minted}
  \caption{Commands supported}
  \label{lst:commands}
\end{listing}

\vspace{-1.5mm}
\subsubsection*{Command Layer}
\label{sec:command-layer}
\vspace{-1.5mm}

This is a software layer that converts requests for data by higher
layers into messages for the sampling layer. High level requests for
data are specified in the form of policies. Each policy includes a
list of locations, the list of sensors and individual sensing
frequencies. The Driver thread (\S\ref{sec:data-layer}) converts these
into the command messages described earlier, replaces the previous
policy with the new policy and keeps the new policy going until the
next policy is received from the higher layer.

When a new sensor joins the list, the default policy is to repeat in a
cycle -- sense for 20 seconds at default frequencies, stop and send
all the data. So the command cycle is -- START (20 seconds), STOP,
SEND. The default frequencies vary depending on the sensor type. For
accelerometer, we set the default frequency to be 20 Hz and for air
quality sensing, 1 Hz. However, the frequency at which data is
reported ultimately depends on the fidelity of the sensor. If the
instructed frequency is higher than the maximum frequency at which the
sensor can sense, then the sensor will sense at its maximum possible
frequency.

\vspace{-1.5mm}
\subsubsection*{ML Layer for Control}
\label{sec:ml-layer}
\vspace{-1.5mm}

This machine learning (ML) layer for control is the layer that
computes application specific policies that need to be implemented
based on observations and constraints. The data collection objective
usually is to collect as much sensor data as necessary to perform the
required analytics and make subsequent decisions. The application
layer constraint is typically to ensure a base sampling requirement to
achieve the urban sensing objective. The lower-layer constraints in
the problem are three-fold: network, cost and power. This layer
constantly accesses the data that is received from the edge nodes,
runs analytics on them, such as field estimation, forecasting, hotspot
detection, etc. and outputs policies, after attempting to satisfy the
constraints. As an example, if the input from the ground are GPS
traces from several mobile phones placed in cars, and the objective is
to detect as many traffic hotspots (road links with high congestion)
as possible, then this layer would first map the location reports into
road traffic segments/links, then estimate average speeds in those
segments based on the movements of the mobile phones, and then
estimate future average speeds based on current and historical
trends. The road segments with very low forecasted average speeds are
potential hotspots. Following this forecasting procedure, a new policy
would be generated, such as to increase frequency of sensing in those
segments of interest and stop sensing from those segments that exhibit
close to free-flow traffic.

\vspace{-2mm}
\section{Applications and Experiences}
\label{sec:applications}
\vspace{-2mm}

In this section, we outline three specific applications that support
different types of commands and policies generated based on sensing
objectives and the constraints. We describe three example objectives
here in order to present our case for the MUSIC platform: (a)
\emph{spatial coverage}, for air quality sensing; (b) \emph{hotspot
  detection}, for traffic congestion inference; (c) \emph{object
  recognition algorithms}, for cleanliness monitoring.

\subsection{Spatial Coverage Mapping of Air Quality in Delhi}
\label{sec:spatial-coverage}
\vspace{-2mm}

We have deployed a air quality monitoring platform in Delhi comprising
a few sensors (highly polluted city) built on top of the MUSIC
platform.  When collecting data such as air quality or GPS locations
from buses on the roads, our spatial coverage objective is to ensure
that the sensors report data is a ``coordinated'' and intelligent
manner so as to avoid excessive data reporting and power
consumption. In our formulation, the spatial coverage mapping function
takes as input sensor readings from multiple locations over a period
of time and computes a mean field for pollution in a locality using
sensor interpolation techniques. The problem, then, at every regular
interval, is to make a decision for each sensor on whether or not it
should sense and at what frequency, so that the constructed field has
as low as error as possible while satisfying both the network and
power constraints.  Using the MUSIC system, this would be achieved by
implementing such a function at the ML Layer. The error computed at
the ML Layer would then be translated into actual commands to be sent
(such as increase sampling frequency) at the Control Layer. Then the
commands would be prepared by the Sampling layer, conforming to the
lower-level message format expected by the Communication Layer.

A simpler implementation of the spatial coverage mapping function and
control for air quality would work as follows. A threshold distance
for good separation between two sensors is about 0.5 km, in the case
of air quality sensors. This is based on real world sensor placements
using our low-cost air quality monitors in Delhi. An example
control logic would work simply as follows -- monitor pairwise
distances of sensors on the ground, and if any pair is closer than 0.5
km, send a STOP command to one of them to stop sensing. Once it
crosses the 0.5 km boundary, send a START command again. This can be
done easily by maintaining state in the backend. Which sensor we
choose to send the command to depends on the current state of that
sensor, including their battery levels.

\begin{figure}
  \centering
  \begin{subfigure}[b]{0.3\linewidth}
    \centering
    \includegraphics[width=\linewidth]{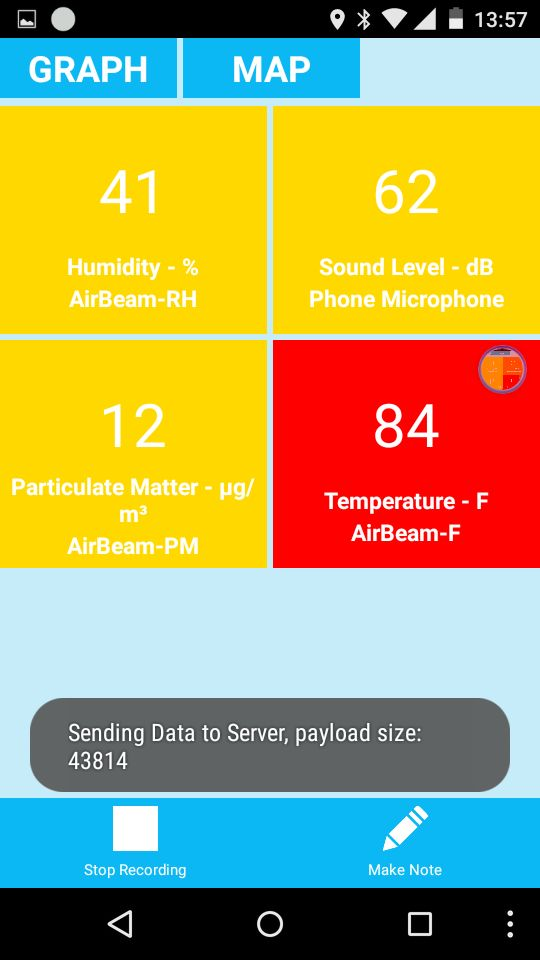}
    \caption{Sending data on SEND command}
    \label{fig:music-app-3}
  \end{subfigure}
  ~
  \begin{subfigure}[b]{0.6\linewidth}
    \centering
    \includegraphics[width=\linewidth]{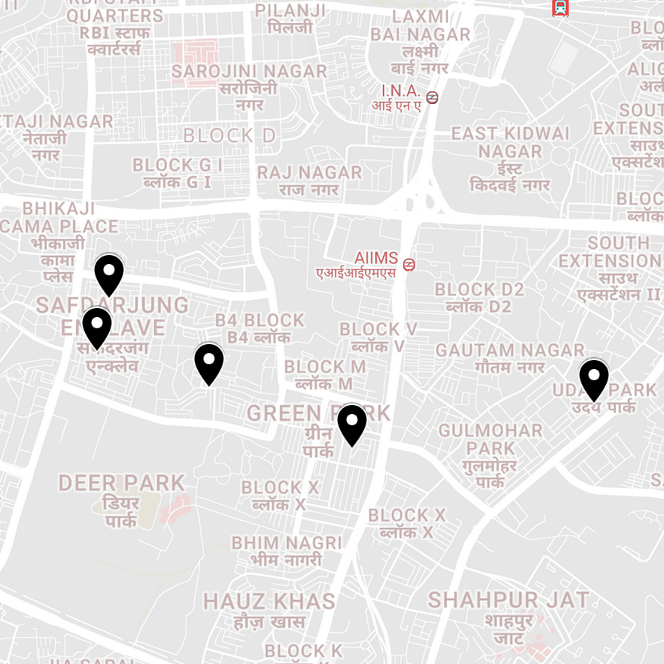}
    \caption{Sensor Locations}
    \label{fig:sensor-locations}
  \end{subfigure}
  \caption{MUSIC app for air quality}
  \label{fig:music-aq}
\end{figure}

We used this system for real world air quality sensing on roads in the
city of Delhi, India. We placed Airbeam air quality sensors
\cite{aircasting} in a small region in the city of Delhi. Each sensor
is hooked up to a mobile phone that runs our smart sensing app. The
default sensing setting is to sense at the maximum frequency possible
by the air quality sensor (1 Hz) for about 20 seconds, and then break
for another 10 seconds and then restart the cycle. So the command
cycle is -- START (20 seconds), STOP, SEND. The server waits till all
the data is received before sending then next START command. This
cycle was able to capture data sufficiently well. Figure
\ref{fig:music-app-3} shows the app in action. We placed 5 such
sensors sensors in 5 different locations in a small region in South
Delhi in India, as shown in figure \ref{fig:sensor-locations}. We aim to 
extend this to a larger scale deployment.

\subsection{Road Traffic Hotspot Detection in New York City}
\label{sec:hotspot}
\vspace{-2mm}


Hotspots are interesting because they point to highly localized points
of unusual activity, which may be a source for a larger problem. For
instance, a highly localized activity such as a road accident or
construction work on a particular road segment would exhibit ripple
effects that result in traffic congestion in farther parts of the
city. In our formulation, the hotspot detection function is a function
that takes as input recent historical readings from an array of
sensors and outputs a boolean array, indicating whether or not each
sensor is located in a hotspot. Assuming that no two sensors in a
neighborhood are co-located, we define a hotspot location as one that
exhibits an unusually higher or lower average reading than the
others.

In the MUSIC system, just as in the case of the spatial coverage
function, the hotspot detection function would be implemented in the
ML Layer. When any hotspots are detected (i.e. if there are 1s in the
boolean array output), then a request for higher frequency data is
sent. This is then translated into appropriate commands by the Control
Layer.
In an application to monitor road traffic congestion, GPS readings are
recorded from mobile vehicles. With an array of location traces (as
(lat,lon) coordinates) along with timestamps from several mobile
vehicles over a period of time, average vehicular traversal speeds can
be computed for each road segment or link. A traffic hotspot may be
defined arbitrarily, but let us say we call it a road segment a
\emph{hotspot} if there is a sudden drop in vehicle speed in that
segment over a sustained period of time. A simple policy to maximize
likelihood of detecting hotspots while being network- and power-aware,
then, is to keep sensing at a certain frequency to begin with, reduce
it as the speed on the segment approaches the free-flow traffic speed,
and increase frequency of sensing whenever there is a dip in the
speed. If the frequency drops below a certain threshold, then the
camera may be activated in order to capture images so that the
situation may be assessed more accurately.

\begin{figure}
  \centering
  \begin{subfigure}{0.5\linewidth}
    \centering
    \includegraphics[width=\linewidth]{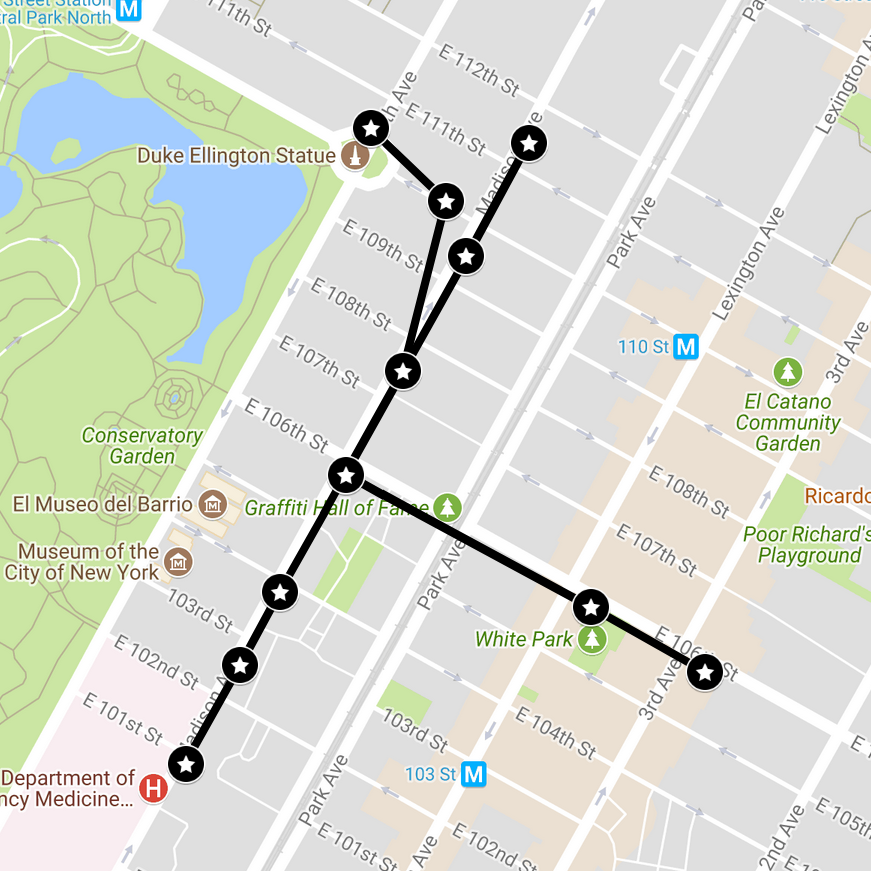}
    \caption{Road segments in Upper East Side in New York City}
    \label{fig:nyc-traffic-segments}
  \end{subfigure}
  ~
  \begin{subfigure}{0.7\linewidth}
    \centering
    \includegraphics[width=\linewidth]{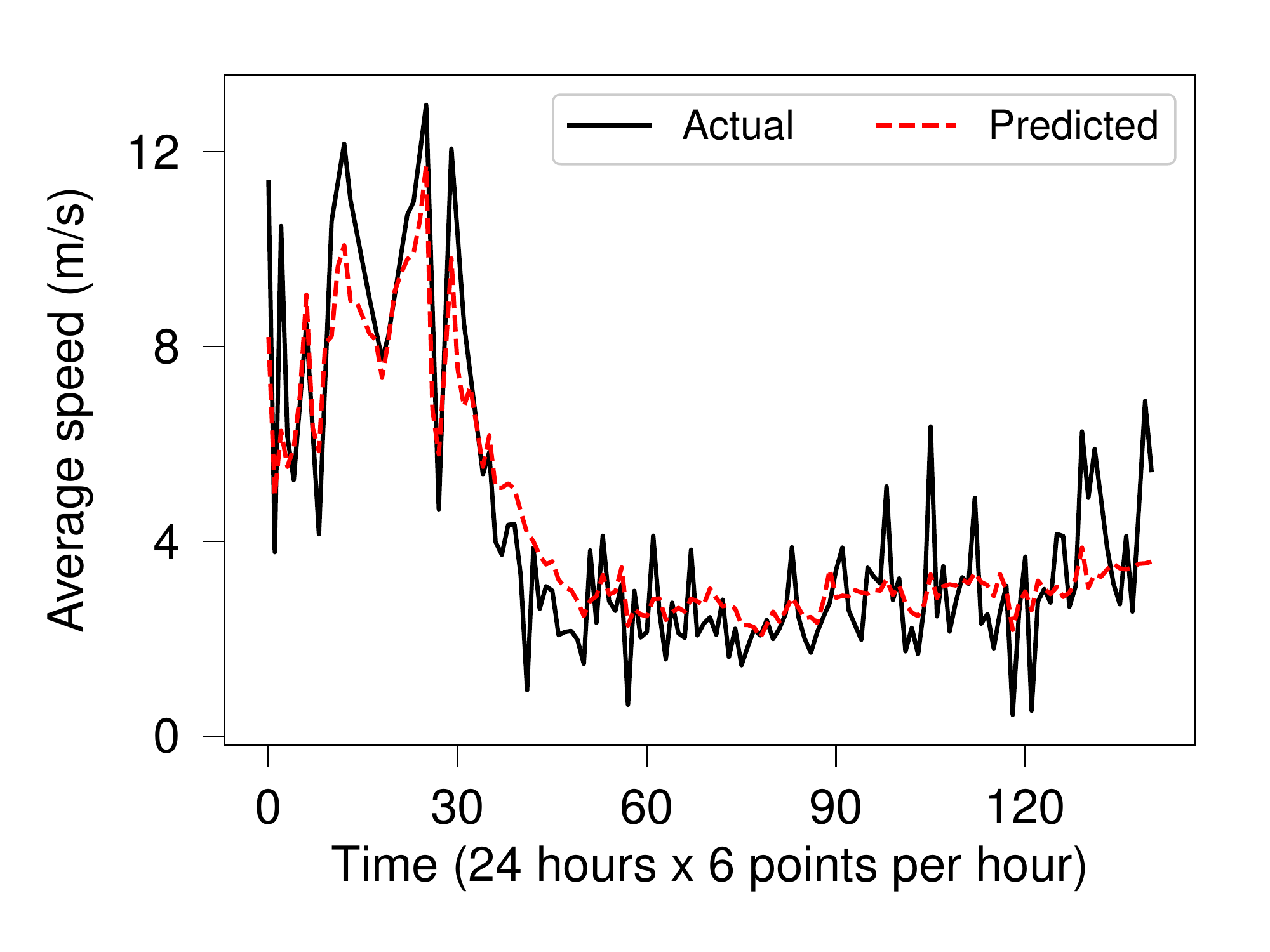}
    \caption{Prediction of average link speeds at future times}
    \label{fig:nyc-traffic-pred}
  \end{subfigure}
  \label{fig:nyc-traffic}
\end{figure}

We have tested a version of our MUSIC platform for traffic congestion
detection with a small number of mobile vehicles. To demonstrate the
potential of this application, we have worked on congestion and
hotspot detection using open GPS traces in New York City, using city
bus mobility trace data from the NYC MTA \cite{nycmta}. The data was
available for three months for all the buses in the city. One can
imagine a setting where all these buses are supported and controlled
by a MUSIC traffic application.  To detect congestion hotspots, we
adopted an approach as follows  -- divide roads into segments defined
by the portion between two consecutive bus stops on any bus route,
determine average bus traverse speed in every segment in every 10
minute interval for 3 months and use the first two-thirds of the
resulting timeseries data to train a predictive model to forecast
average speeds in neighboring segments. We built a graph neural
network to achieve the predictive
task. Figure~\ref{fig:nyc-traffic-pred} shows the speed trend on a
single day on a single segment, with the dotted line showing our model
prediction. Figure~\ref{fig:nyc-traffic-segments} illustrates some
sample segments from the dataset. We have obtained clearance from a transportation board 
to perform a larger scale rollout of the MUSIC traffic application in a large city
in a developing country.

\subsection{Urban Cleanliness Monitoring in Delhi}
\label{sec:cleanliness}
\vspace{-2mm}

Urban spaces (both indoor and outdoor) in many large cities are
dirty. The dirt detection application aims to demonstrate how the
camera of a distributed collection of mobile phones can be used as a
sensor for dirt detection in conjunction with image processing
algorithms at the back-end to detect ``dirty items on the road''.

\begin{figure}
  \centering
  \includegraphics[width=0.5\linewidth]{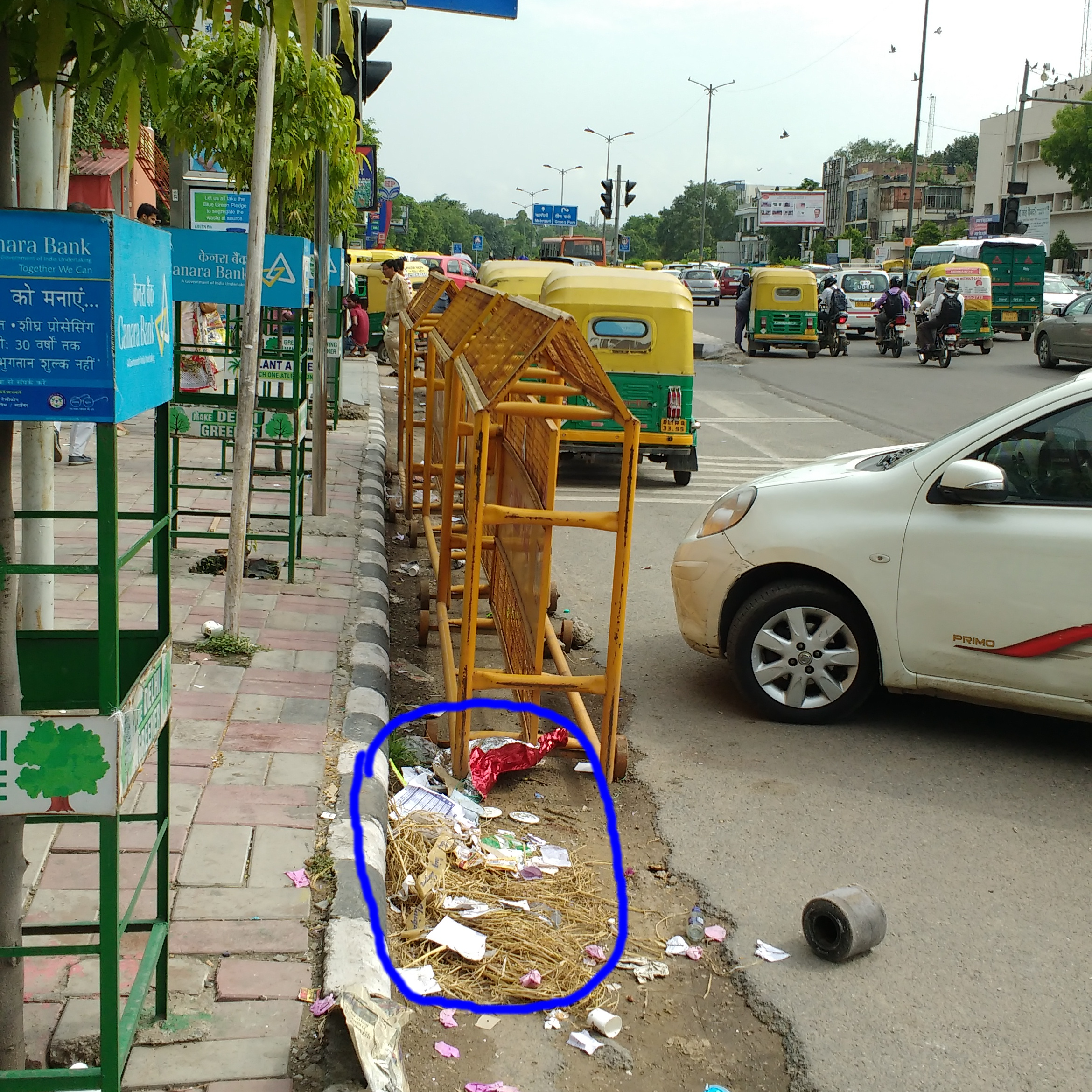}
  \caption{Sidewalk in Delhi showing litter}
  \label{fig:delhi-road}
\end{figure}

We developed a application to detect dirt, dust and indications of
lack of cleanliness in urban spaces. Using deep neural networks, a
prototype was implemented successfully to detect dirt patches in
hospital rooms in the city of Delhi, India. We are currently working
on extending this work for wider cases such as roadscape
photos. Figure~\ref{fig:delhi-road} shows an image taken with our
application, triggered by the CAPTURE\_IMAGE command, with the blue
circle showing the manual annotation for training the system. With our
understanding that such algorithms can be developed feasibly, these
algorithms can then be integrated into our backend so that they be
used for making decisions on sensing in the backend. In the same
example, if more than a threshold amount of dirt is detected, then
sensing of other sensors can be started, such as PM concentration or
dust concentration or humidity and so on. This application is to
primarily demonstrate an alternative setting where the sampling
frequency can be controlled based on the output of the cleanliness
detection algorithms to determine the areas that require to be sampled
more than others. This application has been tested in indoor and
outdoor contexts in Delhi and we plan to deploy it at scale in the
future.

\vspace{-2mm}
\section{Conclusion}
\label{sec:conclusion}
\vspace{-2mm}

In this paper, we have presented MUSIC, a programmable
end-to-end platform for various urban sensing and telemetry
applications.
We presented the MUSIC stack in detail that enables us to implement
the intelligence in the system via centralized control and determining
application specific sensing policies to meet specific urban sensing
objectives subject to network, power and cost constraints. We showed
early experiences of the system for three applications -- spatial
coverage mapping for urban air quality, road traffic congestion
detection and dirt detection in urban spaces.

\vspace{-2mm}
\section*{Acknowledgements}
\label{sec:acks}
\vspace{-2mm}

We acknowledge the contributions of graduate students Zal Bhathena,
Hongtao Cheng, Wesley Painter, Wenliang Zhao and Joseph Zuhusky
(alphabetical order) to the frontend and backend code, which were
implemented as part of their project in the course "Foundations of
Networking". We acknowledge the collaboration with the research team
at Evidence for Policy Design (EPoD) India at IFMR in Delhi for our
urban air quality monitoring experiments. The code for this project
was adapted from that of the AirCasting project \cite{aircasting}
which is open-source until GNU GPL, and the authors thank the
developers.

\appendix

\bibliographystyle{plainnat}
\small{
  \bibliography{refs}
}
\label{last-page}

\end{document}